\documentstyle[11pt,twoside,jltp,epsfig]{article}

\title{On the energy landscape at the glass transition}

\author{Ulrich Buchenau}\address{Institut f\"ur Festk\"orperforschung,
Forschungszentrum J\"ulich, Postfach 1913, D--52425 J\"ulich,
Federal Republic of Germany}

\runninghead{U. Buchenau}{On the energy landscape at the glass
transition}

\begin{document}

\begin{abstract}
A recent hypothesis claims that the glass transition itself,
though it is a very pronounced relaxation peak, is no separate
relaxation process at all, but is just the breakdown of the shear
modulus due to the weak elastic dipole interaction between all the
quasi-independent relaxation centers of the glass. Two derivations
are considered, one of them in terms of a breakdown of the shear
modulus and the second in terms of a divergence of the shear
compliance. Mechanical relaxation data from the literature for
vitreous silica, glycerol, polymethylmethacrylate and polystyrene
are found to be consistent with the first hypothesis.

PACS numbers: 64.70.Pf
\end{abstract}

\maketitle

\vspace{0.3in}

In 1972, Sigi Hunklingers famous experiment \cite{sigi}
demonstrated the two-level nature of the universal low-temperature
excitations in glasses \cite{pohl}. The experiment gave strong
support to the energy landscape concept \cite{goldstein}, in which
the two-level states are explained in terms of tunneling between
adjacent minima of the energy landscape, with a low barrier
between them \cite{phil}.

If the energy landscape concept is indeed the correct explanation,
these two-level states are expected to be a small fraction of a
vast number of local relaxation centers in the glass. In this
view, the relaxation spectrum of the energy landscape spans a wide
range, from the tunneling states at the low-barrier side to the
Johari-Goldstein peak \cite{johari} at the high-barrier side.
Depending on frequency, this Johari-Goldstein peak is either
observed in the glass close to the glass transition temperature
$T_g$ or even above $T_g$ in the supercooled liquid.

While the energy landscape concept in itself seems to be
reasonably well established, the extent to which it can be
represented by an ensemble of independent double-well potentials
is a widely debated question. With respect to the tunneling
states, this question begins to be answered by experiments in the
mK range \cite{sigi2}. Again, Sigi Hunklinger played a central
role in the performance and interpretation of these experiments,
approximately thirty years after establishing his scientific
reputation with the demonstration of two-level states \cite{sigi}.
According to these new experiments, the concept of isolated
tunneling states seems to break down in the mK range. A plausible
explanation is the coupling of the tunneling states by the
interaction of their elastic dipole moments \cite{burin}, which
leads to the formation of coupled pairs of tunneling states.

The present paper deals with the question whether the same elastic
dipole interaction provides the key to the unsolved riddle of the
glass transition. Here, we assume that this is indeed the case. In
particular, we assume that the shear relaxation is due to an
ensemble of double-well potentials which are weakly coupled by the
interaction of their elastic dipole moments. The ensemble
comprises the tunneling states at the low-barrier end as well as
the Johari-Goldstein relaxation at the high-barrier end, possibly
with rather different atomic jump vectors, but all of them
interacting with each other. The reader should be aware that this
is only one of many explanations in the literature
\cite{edi,vogel}.

One can treat the elastic-dipole interaction in two different
mean-field approaches (i) in terms of the shear modulus (ii) in
terms of the shear compliance. Here, equations for both approaches
are derived and compared to dynamical mechanical shear data for
four glassformers at their respective glass temperatures $T_g$.

In the shear modulus treatment \cite{bu}, one considers a small
initial shear deformation $\epsilon$ switched on at the time
$t=0$. The initial shear stress $\sigma=G\epsilon$, where $G$ is
the infinite frequency shear modulus. The shear stress decay with
increasing time is described \cite{ferry} in terms of the
rheological function $H(\tau)$
\begin{equation}\label{gt}
G(t)=G-\int_{-\infty}^\infty H(\tau)(1-{\rm e}^{-t/\tau})d\ln\tau.
\end{equation}

In the energy landscape concept, relaxation occurs via thermally
activated jumps over the barriers $V$ between different energy
minima. The corresponding relaxation time $\tau_V$ obeys the
Arrhenius relation
\begin{equation}\label{arrh}
\tau_V=\tau_0\exp(V/k_BT)
\end{equation}
with $\tau_0\approx 10^{-13}s$. Therefore we replace the
rheological function $H(\tau)$ by the barrier density function
$f(V)$
\begin{equation}\label{rheo}
H(\tau_0{\rm e}^{V/k_BT})=H(\tau_V)=Gk_BTf(V).
\end{equation}

Inserting this definition into equation (\ref{gt}) for the time
dependence of the shear modulus gives
\begin{equation} G(t)=G\left(1-\int_0^\infty f(V)(1-{\rm
e}^{-t/\tau_V})dV\right)\approx G\left(1-\int_0^{V_t}
f(V)dV\right),
\end{equation}
where $V_t$ is the barrier with $\tau_V=t$.

In order to take the influence of the elastic dipole interaction
into account, one assumes a true barrier density $f_0(V)$ of the
energy landscape. At the barrier height $V_t$, the jump between
the two minima will tend to occur at the time $t$, when the shear
stress has decreased to $G(t)\epsilon$. Thus the remaining shear
stress energy is only a fraction $G(t)^2/G^2$ of the initial one.
If we now consider the microscopic situation at the local
relaxation center as unchanged, with the same initial shear
distortion, then the free energy change by the jumps of the
relaxation center remains unchanged. This means that the
relaxation center releases the same amount of stress energy. But
it reduces a stress energy weakened by $G(t)^2/G^2$, so its
effectivity in bringing the shear modulus down to zero is
increased by the reverse of this factor.

Therefore the mean-field approach for the shear modulus reads
\begin{equation}\label{f0}
f_0(V)\equiv f(V)\left[1-\int_0^Vf(v)dv\right]^2.
\end{equation}

In a physical picture, the enhancement of $f_0(V)$ is due to
induced jumps of lower-barrier relaxation centers in the
neighborhood of the given relaxation center with barrier $V$,
which occur quasi-instantaneously after its jump.

The equation (\ref{f0}) has the back-transformation \cite{bu}
\begin{equation}\label{fin}
f(V)=\frac{f_0(V)}{\left[1-3\int_0^{V}f_0(v)dv\right]^{2/3}}.
\end{equation}

In the shear modulus approach, the breakdown of the shear rigidity
occurs when the integral of $f_0(V)$ over $V$ reaches 1/3. The
corresponding barrier is called Maxwell barrier, because it
determines the Maxwell time $\tau_M$ (the shear stress relaxation
time) through the Arrhenius relation, eq. (\ref{arrh}). It is
given by the 1/3-rule derived in ref. \cite{bu}
\begin{equation}\label{third}
\int_0^{V_M}f_0(V)dV=\frac{1}{3}.
\end{equation}

The breakdown occurs in a rather dramatic way, because the
relaxing entities at the critical Maxwell barrier value receive a
strong enhancement, to such an extent that one is tempted to
assume a separate $\alpha$-process which has nothing to do with
the secondary glass relaxations. In fact, this more or less
unconscious assumption underlies most of the present attempts to
understand the glass transition \cite{edi}. The above treatment
shows such an assumption to be unnecessary; what one sees at the
glass transition are simple Arrhenius relaxations of no
particularly large number density $f_0(V)$, blown up to impressive
size in $f(V)$ by the small denominator of eq. (\ref{fin}).

The shear modulus approach can be tested by plotting the product
$G''G'^2$ of experimental dynamical shear data as a function of
frequency or temperature. In fact, the true barrier density
$f_0(V)$ can be calculated approximately from the equation
\begin{equation}\label{f0vm}
V_Mf_0(k_BT\ln\frac{1}{\omega\tau_0})\approx\frac{2}{\pi}
\frac{G''G'^2}{G^3}\ln\frac{\tau_M}{\tau_0},
\end{equation}
where $G$ is again the infinite frequency shear modulus.

One can build the analogue of the shear modulus approach, starting
from the shear compliance. In the shear compliance treatment, one
considers a small initial shear stress $\sigma$ switched on at the
time $t=0$. The initial shear distortion $\epsilon=J\sigma$, where
$J=1/G$ is the infinite frequency shear compliance. The shear
distortion increase with increasing time is described \cite{ferry}
in terms of the rheological function $L(\tau)$
\begin{equation}\label{jt}
J(t)=\frac{1}{G}+\int_{-\infty}^\infty L(\tau) (1-{\rm
e}^{-t/\tau})d\ln\tau+\frac{t}{\eta_0},
\end{equation}
where $\eta$ is the viscosity. In the following, we omit this
viscosity term, because we are only interested in the breakdown of
the shear rigidity.

In the energy landscape concept, we replace the rheological
function $L(\tau)$ by another barrier density function $l(V)$
defined by
\begin{equation}
L(\tau_0{\rm e}^{V/k_BT})=L(\tau_V)=\frac{k_BTl(V)}{G}.
\end{equation}

Inserting this definition into equation (\ref{jt}) for the time
dependence of the shear compliance gives
\begin{equation}\label{jtl} J(t)=\frac{1}{G}\left[1+\int_0^\infty l(V)(1-{\rm
e}^{-t/\tau_V})dV\right]\approx\frac{1}{G}\left[1+\int_0^{V_t}
l(V)dV\right].
\end{equation}
where again $V_t$ is the barrier with $\tau_V=t$.

Again, the influence of the elastic dipole interaction is taken
into account assuming a true barrier density $l_0(V)$ of the
energy landscape. But now, the situation at the barrier height
$V_t$ for the time $t$ is different. The shear stress is still the
initial one, but the shear distortion has increased by $J(t)G$.
Therefore the free energy change by the jumps of the relaxation
center is increased by $J(t)^2G^2$.

Thus the mean-field approach for the shear compliance reads
\begin{equation}\label{l0}
l_0(V)\equiv\frac{l(V)}{\left[1+\int_0^Vl(v)dv\right]^2}.
\end{equation}

Its back-transformation is
\begin{equation}\label{finl}
l(V)=\frac{l_0(V)}{\left[1-\int_0^{V}l_0(v)dv\right]^2}.
\end{equation}

We see that the results for the shear compliance approach (ii)
differ markedly from the shear modulus approach (i) proposed
earlier \cite{bu}. The Maxwell barrier is now given by the
divergence of the compliance
\begin{equation}\label{lint}
\int_0^{V_M}l_0(V)dV=1.
\end{equation}

Again, the relaxing entities at the critical Maxwell barrier value
receive a strong enhancement. The shear compliance approach can be
tested by plotting the ratio $J''/J'^2$ of experimental dynamical
shear data as a function of frequency or temperature. The true
barrier density $l_0(V)$ can be calculated approximately from the
equation
\begin{equation}\label{l0vm}
V_Ml_0(k_BT\ln\frac{1}{\omega\tau_0})\approx\frac{2}{\pi}
J''J/J'^2\ln\frac{\tau_M}{\tau_0}.
\end{equation}
Since the influence of the viscosity was neglected, this equation
is only expected to hold down to the frequency with
$\omega\tau_M=10$, at least for non-polymeric glass formers. This
is a drawback of the compliance approach. If one tries to include
the viscosity, the equations get complicated and rather difficult
to handle.

In any case, one is now able to compare dynamical mechanical shear
data from experiment to the two theoretical approaches. Fig. 1
begins with the shear modulus approach, showing $G''G'^2$-values
from measurements at the glass transition. The comparison is done
for the four glass formers $SiO_2$, glycerol, polystyrene and
polymethylmethacrylate. Also included is the simplest theoretical
case, with $f_0(V)=constant=1/3V_M$, the so-called {\it generic
case}. The figure is scaled in such a way that this generic case
is 1. The scaling requires the knowledge of the Maxwell time
$\tau_M$. For silica and glycerol, the classical Maxwell relation
$\tau_M=\eta/G$ was used. For the two polymers, $\tau_M$ was taken
from the condition $\omega_{max}\tau_M=1$, where $\omega_{max}$
was the maximum of the peak in $G''$. In order to cover five
decades in such a mechanical measurement, one usually needs to do
measurements at several temperatures around $T_g$ and use the
time-temperature scaling to obtain the full curve.

The scaling $G''G'/G^3$ also requires the knowledge of the
infinite frequency shear modulus $G$. Therefore one needs not only
a measurement of $G'$ and $G''$, but also a Brillouin measurement
of the transverse sound wave velocity $v_t$ at $T_g$, to determine
$G$ via $G=\rho v_t^2$, where $\rho$ is the density at $T_g$.
Table I summarizes the $T_g$ and $G$ values of the four glass
formers, together with the relevant references. The glass
temperature for glycerol is somewhat higher than the usual value,
because the master curves for $G'$ and $G''$ were given
\cite{donth} for this temperature.

\bigskip
\begin{center}
Table I: Glass transition data.
\end{center}
\begin{center}
\begin{tabular}{|l|c|c|c|c|}
\hline     substance       & SiO$_2$ & glycerol &   polystyrene &
polymethylmethacrylate
\\ \hline      $T_g$ (K)       &    1449     & 192.5 &  363  & 383      \\
    $G(T_g)$ (GPa)    &  33.8 &  4.3  &  1.5     &   1.86        \\
    $\eta(T_g)$ (GPa s) & 2512 & 10  &      &     \\
    $\tau_M$  (s)  &  74.3  &  2.3   &  100   &  100    \\
    ref. $G$ &\cite{dardy}&\cite{fio}&\cite{strube}&\cite{kruger}\\
    ref. $G'$, $G''$&\cite{mills}&\cite{donth}&\cite{beiner}&\cite{perez}\\
    $m$  & 20 & 53 & 138 & 145  \\
      \hline
\end{tabular}
\bigskip
\end{center}

\begin{figure}
\centerline{\epsfig{file=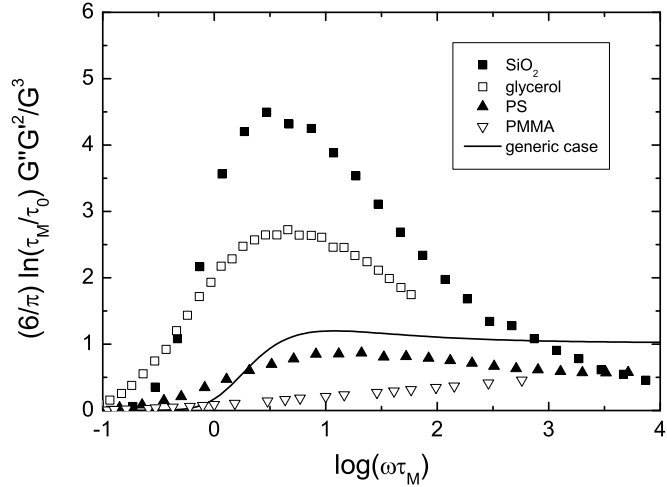,height=3in,angle=0}}
%
\caption{Comparison of data measured at the glass transition of
$SiO_2$, glycerol, polystyrene (PS) and polymethylmethacrylate
(PMMA) to the generic case $f_0(V)=1/3V_M$ in the shear modulus
approach.} \label{fig:sigi1}
\end{figure}

In Fig. 1, all four experimental curves show a cutoff at
$\omega\tau_M=1$. If one looks closely, one observes that this
experimental cutoff is slightly broader than in the theoretical
calculation for the generic case, where a sharp cutoff of $f_0(V)$
at $V_M$ was assumed. The two polymers lie decidedly lower than
the generic case, the two other cases lie higher.

The corresponding comparison to the compliance approach in Fig. 2
suffers from the fact that for silica and glycerol, it is only
meaningful above $\omega\tau_M=10$, because of the neglected
viscosity. Again, silica shows the strongest rise towards the
value $\omega\tau_M=1$, PMMA the weakest.

But even for the two polymers, which have such a high viscosity
that it can be indeed neglected, the compliance approach does not
work well; there remains a pronounced peak in the near
neighbourhood of $\omega\tau_M=1$, much more pronounced than in
the shear modulus approach of Fig. 1. Therefore we conclude that
the shear modulus approach is the better description.

This is not unexpected: The shear modulus approach corresponds to
a generalized Maxwell model, with an infinite number of parallel
Maxwell elements, while the shear compliance approach corresponds
to an infinite number of Voigt elements in series \cite{ferry}.
The description of an ensemble of localized relaxation centers in
a threedimensional viscoelastic continuum in terms of parallel
elements is probably more suitable.

Further support for this conclusion is supplied by the obvious
connection between the values of the fragility $m$ in Table I
(taken from ref. \cite{bohmer}) and the values of $V_Mf_0(V_M)$
extrapolated from Fig. 1. The fragility $m=\partial\log
\eta/\partial(T_g/T)$ is defined in terms of the steep rise of the
viscosity $\eta$ towards the glass temperature $T_g$ with
decreasing temperature in the supercooled liquid. Silica has the
lowest fragility of all known glass formers, while the two
polymers belong to the very high fragility end of the scale. A
high fragility means a strong decrease of the Maxwell barrier
$V_M$ with increasing temperature. Fig. 1 indicates that a high
$m$ seems to be related to a low $V_Mf_0(V_M)$, and viceversa.

\begin{figure}
\centerline{\epsfig{file=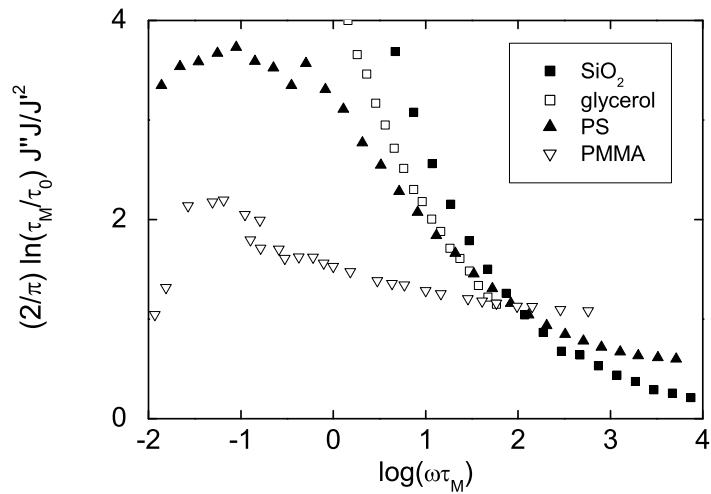,height=3in}}
%
\caption{Comparison of data measured at the glass transition of
$SiO_2$, glycerol, polystyrene (PS) and polymethylmethacrylate
(PMMA) in the shear compliance approach.} \label{fig:sigi2}
\end{figure}

From the 1/3-rule, eq. (\ref{third}), such a behaviour is
expected: If $f_0(V)$ tends to increase with increasing
temperature, a low value of $f_0(V_M)$ means a fast decrease of
$V_M$ with increasing temperature. Thus one gets at least a
qualitative understanding of the old fragility riddle \cite{edi}:
With increasing temperature, more and more minima of the energy
landscape get populated. This leads to an increase of $f_0(V)$,
which in turn leads to a decrease of $V_M$. This decrease is more
dramatic for a low value of $V_Mf_0(V_M)$ than for a high one.

Of course, Fig. 1 does not prove the shear modulus description of
the glass transition \cite{bu} beyond any possible doubt. But the
plot of Fig. 1 is a new, quantitative and probably meaningful way
to study the relation between the primary relaxation (the shear
stress relaxation to the value zero) and the secondary relaxations
at higher frequencies.

\end{document}